\documentclass[twocolumn,prl,showpacs,floatfix,amsmath,amsfonts]{revtex4-1}
\usepackage[latin9]{inputenc}
\usepackage{verbatim}
\usepackage{amsmath}
\usepackage{amssymb}
\usepackage{graphicx}
\usepackage{dcolumn}
\usepackage{hyperref}
\pdfoutput=1

\makeatother

\begin{document}
\title{Excitation of localized graphene plasmons by a metallic  slit }
\author{Yu. V. Bludov$^{1}$\email{bludov@fisica.uminho.pt}, N. M. R. Peres$^{1,2}$, and
M. I. Vasilevskiy$^{1,2}$}
\affiliation{$^{1}$Department of Physics, Center of Physics, and QuantaLab, University
of Minho, Campus of Gualtar, 4710-057, Braga, Portugal}
\affiliation{$^{2}$International Iberian Nanotechnology Laboratory
(INL), Av. Mestre Jos\'{e} Veiga, 4715-330 Braga, Portugal}
\begin{abstract}
In this paper we show that graphene surface plasmons can be excited when an electromagnetic wave packet impinges on a single metal slit covered with graphene. The excitation of the plasmons localized over the slit is revealed by characteristic peaks in the absorption spectrum. It is shown that the position of the peaks can be tuned either by the graphene doping level or by the dielectric function of the material filling the slit. The whole system forms the basis for a plasmonic sensor when the slit is filled with an analyte. 
\end{abstract}
\maketitle

\section{Introduction}

The diffraction of electromagnetic (EM) waves on metallic structures
gives rise to a series of interesting phenomena, such as the Wood and Rayleigh
anomalies \cite{difr-slits-Wood1902-pm,difr-slits-Rayleigh1907-prsa}
and the extraordinary optical transmission \cite{spp-eot-Ebessen1998-nat}.
Theoretical models for these phenomena have been elaborated 
by modelling the metal as perfect electric conductor (PEC), as in
Refs. \cite{difr-slits-Porto1999-prl,difr-slits-Sturman2011-prb,grating-Maradudin2016-jopt}. 
One of the fundamental problems in the nano-optics is the diffraction of light
on a single slit of subwavelength width perforated in a metal.
This kind of diffraction is accompanied by a series of effects such
as funneling of the EM energy into the slit \cite{difr-slit-Sturman2010-prb,difr-slit-Li2018-plasmonics} and field enhancement inside it \cite{difr-slit-Sturman2010-prb},
Fabri-P\'erot resonances across the metallic film \cite{difr-slit-Takakura2001-prl,difr-slit-exp-Yang2002-prl,difr-slit-Bravo2004-pre}, transmittance oscillations with an incidence angle variation in the geometrical optics  limit \cite{difr-slit-Park1994-ieeetap} and its absence in the subwavelength limit\cite{difr-slit-Bravo2004-pre}, and the sensitivity
of the transmittance resonance frequencies to the refractive index
of the material inside the slit \cite{difr-slit-Bravo2004-pre}.

Consideration of the light diffraction on a single slit in real metals \cite{difr-slit-SPP-exp-Isaak2008-prb,difr-slit-SPP-Chang2015-josabI,difr-slit-SPP-Chang2015-josabII},
with surface capable to supporting surface plasmon-polaritons (SPPs)
enriches considerably the physics of the diffraction phenomena \cite{spp-rev-Barnes2003-nat}.
Nevertheless, SPPs in noble metals suffer from relatively high losses
in the visible light wavelength, which considerably shorten their mean
free path. One of the perspective ways to overcome this difficulty
is to use graphene plasmons that can be combined with other materials in order to modify SPP's properties \cite{spp-gr-rev-Bludov2013-ijmfb,spp-gr-rev-deAbajo2014-ACSPhot,spp-gr-rev-Low2014-ACSNano,spp-gr-rev-Xiao2016-fp,spp-gr-rev-Chen2017-nanophot,spp-gr-rev-Luo2013-ssc}. 

As is well known, SPPs in graphene possess both large lifetime and high degree of field confinement \cite{spp-gr-Nikitin2011-prb,spp-gr-rev-Koppens2011-nl}.
This property implies the advantage of using some kind
of hybrid metal-graphene structures, where graphene sustains the propagation
of SPPs, while PEC modifies their dispersion properties. For example,
screening of graphene SPPs by a perfect \cite{spp-gr-acoustic-Principi2011-ssc,spp-gr-acoustic-Gu2013-apl,spp-gr-acoustic-exp-Alonso20016-nat}
or dispersive (Drude) \cite{spp-gr-acoustic-Polini2018-prb} metal adjacent to
it leads to the formation of acoustic SPPs with linear spectrum. Moreover, in such kind of structures SPP's group velocity is quite low compared to polaritons in graphene on a thick dielectric substrate since high wavevectors correspond to relatively low frequencies in the SPP dispersion relation. The latter means that graphene's conductivity
exhibits its nonlocal properties in the THz frequency range and, therefore, gives rise
to the nonlocal SPPs \cite{spp-gr-nonlocal-Horing2016,spp-gr-nonlocal-exp-Alcaraz2018-science,spp-gr-nonlocal-Dias2018-prb,spp-gr-nonlocal-exp-Koppens2017-science}.
Simultaneously, graphene's conductivity (and, consequently, the dispersion properties of SPPs) can be effectively controlled by changing the applied gate voltage \cite{gr-cond-exp-Li2008-natp}, which allows one to achieve the
dynamical tunability of the resonant frequency in the graphene-based
structures \cite{spp-gr-tunable-exp-Ju2011-nn,spp-gr-tunable-Yao2014-nl,spp-gr-tunable-exp-Fei2012-nat,spp-gr-tunable-exp-Chen2012-nat,spp-gr-tunable-exp-Alonso2014-nat}.
Being combined with a metallic grating, variation of gate voltage permits to control spoof plasmons \cite{spoof-Dias2017-ACSphot}.

Since the dispersion properties of SPPs are extremely sensitive to the
dielectric constant of the surrounding medium, plasmonics structures
are widely used for molecular and biosensing \cite{spp-sensor-rev-Anker2008-natmat}.
The use of graphene in plasmonic biosensors \cite{spp-gr-sensor-Huang2017-pssa,spp-gr-sensor-Meshginqalam2017-plasmonics} has an additional advantage, since the above-mentioned tunability of the plasmonic resonance frequency allows for achive it in the spectral range where the strength of the characteristic molecular signal is the
highest \cite{spp-gr-sensor-tunable-Hu2016-ncom}.

Monochromatic plane waves are idealizations never realized in practice. The electromagnetic wave that impinges on the plasmonic structure and couples to SPPs (if the necessary conditions are fulfilled \cite{spp-gr-rev-Bludov2013-ijmfb}), generally is a wave packet, i.e. a superposition of plane waves with close but unequal frequencies and wavevectors, which may represent either a pulse or a focused beam \cite{Born-Wolf}.
In the present paper, we consider the diffraction of a localized
wavepacket on the single rectangular slit in PEC film, which is covered by a graphene sheet encapsulated (that is, cladded) by two h-BN layers at one
side and open on the other side (see Fig. 1). We demonstrate that the electromagnetic wave, when diffracted by the slit edges, excites a standing
wave of SPPs in graphene at a series of resonant frequencies, which
are determined by the graphene doping. At these resonance frequencies,
the slit width contains an integer number of SPP wavelengths. Excitation
of the polaritons yields a series of absorption peaks in the spectrum and these resonant frequencies turn out to be very sensitive to the dielectric constant of the dielectric material filling the slit, as it will be demostrated by our calculated results.

\section{Problem statement and main equations}

\subsection{The structure}

We consider a graphene monolayer, located at
plane $z=0$ and deposited on top of a h-BN layer of thickness $d_{\mathrm{BN}}$ that occupies the spatial domain $0<z<d_{\mathrm{BN}}$. The graphene sheet is covered by another h-BN layer of the same thickness ($-d_{\mathrm{BN}}<z<0$). The lower h-BN
layer is deposited on top of the PEC film of thickness
$d$ (see Fig.1) with the surfaces located at $z=d_{\mathrm{BN}}$ and
$z=d_{\mathrm{BN}}+d$. The PEC film contains the single rectangular
slit of width $W$ ($-W/2<x<W/2$), filled with a dielectric
with the permittivity $\epsilon$. We consider the incident wave packet propagating in the positive $z$-axis direction, localized in the $x$ direction and impinging on the uppermost h-BN layer at normal incidence. Furthermore, it is essential that the wave packet is described by an even amplitude function of $x$ with respect to the vertical symmetry plane $x=0$.  To simplify the calculations, we shall consider the wave packet of a constant amplitude within a certain range of $k_x$, $\vert k_x\vert \leq k_c$, where $k_c \leq \omega /c$, $\omega$ is the frequency and $c$ is the velocity of light in vacuum.

\subsection{Maxwell equations and their solutions}

Since the system under consideration is homogeneous in the direction
$y$ (i.e., $\partial/\partial y\equiv0$), the Maxwell equations
can be decoupled into two subsystems, which govern TM- and TE-polarized
waves. In the following we will consider the case of TM-polarized
waves only, which have the field components 
$\mathbf{E}=\left(E_{x},0,E_{z}\right)$ and
$\mathbf{H}=\left(0,H_{y},0\right)$. Assuming the EM field
time-dependence as $\mathbf{E},\mathbf{H}\sim\exp{\left (-i\omega t\right )}$, we
represent the Maxwell equations for the TM-polarized wave as 
\begin{figure}
\includegraphics[width=8.5cm]{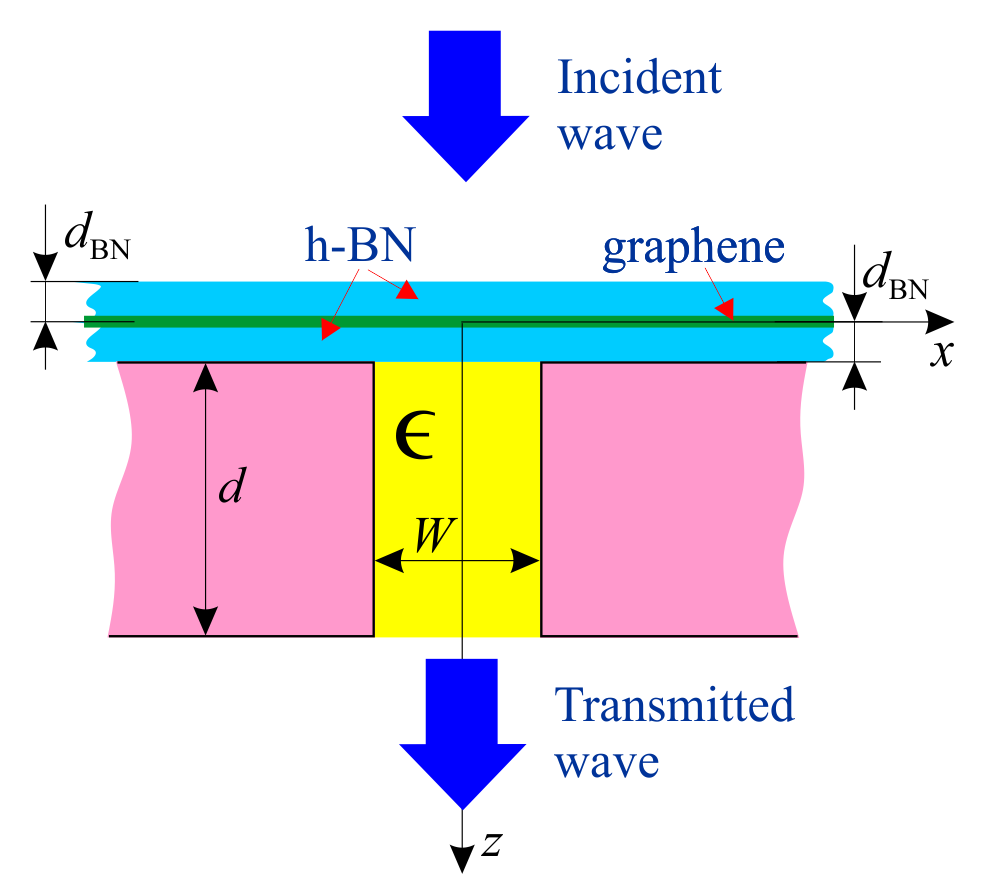}

\caption{Single slit (yellow) in the metal film (pink), covered with h-BN-encapsulated graphene layer. Also shown are the coordinate axes.}

\end{figure}
\begin{eqnarray}
\begin{aligned}
&\frac{\partial E_{x}^{(m)}}{\partial z}-\frac{\partial E_{z}^{(m)}}{\partial x}=\frac{i\omega}{c}H_{y}^{(m)},\label{eq:max_Hy}\\
-&\frac{\partial H_{y}^{(m)}}{\partial z}=-\frac{i\omega}{c}\varepsilon_{xx}^{(m)}E_{x}^{(m)},\label{eq:max_Ex}\\
&\frac{\partial H_{y}^{(m)}}{\partial x}=-\frac{i\omega}{c}\varepsilon_{zz}^{(m)}E_{z}^{(m)}.\label{eq:max_Ez}
\end{aligned}
\end{eqnarray}
The superscripts $m=1,2,3,4,5$ correspond to the
spatial domains $z<-d_{\mathrm{BN}}$, $-d_{\mathrm{BN}}<z<0$, $0<z<d_{\mathrm{BN}}$,
$d_{\mathrm{BN}}<z<d_{\mathrm{BN}}\text{+}d$ and $z>d_{\mathrm{BN}}\text{+}d$,
respectively. The reason for such separation of the whole space into
domains is that inside each domain the dielectric permittivity
is homogeneous and generally is described by the diagonal tensor, in the coordinate frame of Fig. 1:
\begin{eqnarray*}
\hat{\varepsilon}^{(m)}=\left(\begin{array}{ccc}
\varepsilon_{xx}^{(m)} & 0 & 0\\
0 & \varepsilon_{xx}^{(m)} & 0\\
0 & 0 & \varepsilon_{zz}^{(m)}
\end{array}\right).
\end{eqnarray*}
In fact, in the isotropic media the tensor components are the same, for $m=1$ and $m=5$ (vacuum) equal to unity, $\varepsilon_{xx}^{(1)}=\varepsilon_{zz}^{(1)}=\varepsilon_{xx}^{(5)}=\varepsilon_{zz}^{(5)}=1$,
and for $m=4$ (isotropic dielectric) $\varepsilon_{xx}^{(4)}=\varepsilon_{zz}^{(4)}=\epsilon$.
The hexagonal boron nitride (spatial domains $m=2,3$) is a uniaxial medium with unequal tensor components in plane and out of plane, given by \cite{spp-hbn-Kumar2015-nl}: 
\begin{eqnarray}
\begin{aligned}
&\varepsilon_{xx}^{(2,3)}=4.87\left(1+\frac{1610^{2}-1370^{2}}{1370^{2}-\omega^{2}-5i\omega}\right),\\
&\varepsilon_{zz}^{(2,3)}=2.95\left(1+\frac{830^{2}-780^{2}}{780^{2}-\omega^{2}-4i\omega}\right).
\end{aligned}
\end{eqnarray}
Their depencence upon $\omega $ (in cm$^{-1}$) is due to the polar optical phonon response and the numbers entering the above relations are the longitudinal and transverse  optical phonon frequencies and the corresponding damping parameters; the multiplicative factors are the high-frequency dielectric constants. 

Moreover, since both the geometry of the structure and the considered
wave packet are symmetric with respect to the plane $x=0$, we can
seek the solution of the Maxwell equations \eqref{eq:max_Hy}
in the form of Fourier integrals over $k_x >0$,
\begin{eqnarray}
\begin{aligned}
H_{y}^{(m)}(x,z)=\intop_{0}^{\infty}dk_{x}\thinspace h_{y}^{(m)}\left(k_{x},z\right)\cos\left(k_{x}x\right),\label{eq:hym}\\
E_{x}^{(m)}(x,z)=\intop_{0}^{\infty}dk_{x}\thinspace e_{x}^{(m)}\left(k_{x},z\right)\cos\left(k_{x}x\right), 
\end{aligned}
\end{eqnarray}
where $h_{y}^{(m)}\left(k_{x},z\right)$ and $e_{x}^{(m)}\left(k_{x},z\right)$
are the amplitudes of the $k_{x}$-th harmonics of the tangential
components of the magnetic and electric fields, respectively. 

In the semi-infinite medium $m=1$, the solution
of Maxwell equation can be represented in the matrix form as
\begin{eqnarray}
\begin{aligned}
&\left(\begin{array}{c}
h_{y}^{(1)}(k_{x},z)\\
e_{x}^{(1)}(k_{x},z)
\end{array}\right)=\label{eq:HE1-z-from-IR}\\
&=\hat{F}_{k_{x}}^{(1)}\cdot \left(\begin{array}{c}
{\cal H}_{y}^{(i)}\Theta\left(k_{c}-k_{x}\right)\exp\left[ip^{\left(1\right)}\left(k_{x}\right)(z+d_{\mathrm{BN}})\right]\\
h_{r}\left(k_{x}\right)\exp\left[-ip^{(1)}\left(k_{x}\right)(z+d_{\mathrm{BN}})\right]
\end{array}\right).
\end{aligned}
\end{eqnarray}
Here
\begin{eqnarray*}
\hat{F}_{k_{x}}^{(1)}=\left(\begin{array}{cc}
1 & 1\\
\frac{c}{\omega}p^{(1)}\left(k_{x}\right) & -\frac{c}{\omega}p^{(1)}\left(k_{x}\right)
\end{array}\right)
\end{eqnarray*}
is the field matrix, $p^{(1)}\left(k_{x}\right)=\sqrt{\left(\omega/c\right)^{2}-k_{x}^{2}}$
is the wavevector's $z$-component, and $\Theta\left(k_{c}-k_{x}\right)$
is the Heavyside function.  For each line in Eq.\,\eqref{eq:HE1-z-from-IR}, if written explicitly, the first term stands for the incident wavepacket with the amplitude ${\cal H}_{y}^{(i)}$ and the cutoff in-plane wavevector $k_{c}$, which propagates in the positive direction of $z$-axis. Owing to this restriction, all waves constituting the incident wave packet have real $z$-components of the wavevector. The second
term for each line in Eq.\,\eqref{eq:HE1-z-from-IR} is the reflected wave, whose
harmonics {[}with amplitudes $h_{r}\left(k_{x}\right)${]} can be
either propagating in the negative direction of $z$-axis (when 
$k_{x}<\omega/c$) or evanescent (when $k_{x}>\omega/c$) waves. Notice that the sign in the exponent for the evanescent waves is chosen to exclude the harmonics which grows at $z\rightarrow -\infty$.
At the same time, in the other semi-infinite spatial domain, $z>d+d_{\mathrm{BN}}$
($m=5$), the spectrum consist of transmitted waves only, with $p^{(5)}\left(k_{x}\right)\equiv p^{(1)}\left(k_{x}\right)${]},
\begin{eqnarray}
\begin{aligned}
&\left(\begin{array}{c}
h_{y}^{(5)}(k_{x},z)\\
e_{x}^{(5)}(k_{x},z)
\end{array}\right)=\left(\begin{array}{c}
1\\
\frac{c}{\omega}p^{(5)}\left(k_{x}\right)
\end{array}\right)h_{t}\left(k_{x}\right)\times\label{eq:HE5-z}\\
&\times\exp\left[ip^{(5)}\left(k_{x}\right)(z-d_{\mathrm{BN}}-d)\right].
\end{aligned}
\end{eqnarray}
Again, these transmitted harmonics with amplitudes $h_{t}\left(k_{x}\right)$
can be either propagating in the positive direction of the $z$-axis
or evanescent, decaying for $z\rightarrow +\infty$.

In the finite spatial domain $-d_{\mathrm{BN}}<z<0$ (medium $m=2$)
the electromagnetic fields will be represented by means of the transfer
matrix, \begin{widetext}
\begin{eqnarray*}
\hat{Q}_{k_{x},z}^{(2)}=\left(\begin{array}{cc}
\cos\left[p^{(2)}\left(k_{x}\right)\left(z+d_{\mathrm{BN}}\right)\right] & \frac{i\omega}{c}\frac{\varepsilon_{xx}^{(2)}}{p^{(2)}\left(k_{x}\right)}\sin\left[p^{(2)}\left(k_{x}\right)\left(z+d_{\mathrm{BN}}\right)\right]\\
\frac{ic}{\omega}\frac{p^{(2)}\left(k_{x}\right)}{\varepsilon_{xx}^{(2)}}\sin\left[p^{(2)}\left(k_{x}\right)\left(z+d_{\mathrm{BN}}\right)\right] & \cos\left[p^{(2)}\left(k_{x}\right)\left(z+d_{\mathrm{BN}}\right)\right]
\end{array}\right),
\end{eqnarray*}
\end{widetext}as
\begin{eqnarray}
\left(\begin{array}{c}
h_{y}^{(2)}\left(k_{x},z\right)\\
e_{x}^{(2)}\left(k_{x},z\right)
\end{array}\right)=\hat{Q}_{k_{x},z}^{(2)}\left(\begin{array}{c}
h_{y}^{(2)}\left(k_{x},-d_{\mathrm{BN}}\right)\\
e_{x}^{(2)}\left(k_{x},-d_{\mathrm{BN}}\right)
\end{array}\right).\label{eq:HE2-z}
\end{eqnarray}
Here $p^{(2)}\left(k_{x}\right)=\sqrt{\left(\omega/c\right)^{2}\varepsilon_{xx}^{(2)}-k_{x}^{2}\left(\varepsilon_{xx}^{(2)}/\varepsilon_{zz}^{(2)}\right)}$, which is the effective $z$-component of the wavevector in a uniaxial medium \cite {DumelowParkerFarIRPhononsPlasmons}.
In Eq.\,\eqref{eq:HE2-z}, we represented the fields in the hBN substrate
using values $h_{y}^{(2)}\left(k_{x},-d_{\mathrm{BN}}\right)$ and
$e_{x}^{(2)}\left(k_{x},-d_{\mathrm{BN}}\right)$ {[}the EM
field tangential components at $z=-d_{\mathrm{BN}}${]} as free parameters.
This situation is distinct from the case of semi-infinite vacuum {[}see
Eq.\,\eqref{eq:HE1-z-from-IR}{]}, where the amplitudes of the reflected
waves were used as free parameters. These free parameters will be eliminated by matching the fields at the interfaces.

In the medium $m=3$ (spatal domain
$0<z<d_{\mathrm{BN}}$) the field structure is similar to that of
Eq.\,\eqref{eq:HE2-z} with the following replacement: $p^{(2)}\to p^{(3)}$ (as a matter of fact, they are the same, i.e. $p^{(3)}=p^{(2)}$ ),  $\varepsilon_{xx}^{(2)}\to\varepsilon_{xx}^{(3)}$,
$\varepsilon_{zz}^{(2)}\to\varepsilon_{zz}^{(3)}$, and
\begin{eqnarray*}
\left(\begin{array}{c}
h_{y}^{(2)}\left(k_{x},-d_{\mathrm{BN}}\right)\\
e_{x}^{(2)}\left(k_{x},-d_{\mathrm{BN}}\right)
\end{array}\right)\to\left(\begin{array}{c}
h_{y}^{(3)}\left(k_{x},0\right)\\
e_{x}^{(3)}\left(k_{x},0\right)
\end{array}\right).
\end{eqnarray*}
In other words, 
\begin{eqnarray}
\left(\begin{array}{c}
h_{y}^{(3)}\left(k_{x},z\right)\\
e_{x}^{(3)}\left(k_{x},z\right)
\end{array}\right)=\hat{Q}_{k_{x},z}^{(3)}\left(\begin{array}{c}
h_{y}^{(3)}\left(k_{x},0\right)\\
e_{x}^{(3)}\left(k_{x},0\right)
\end{array}\right),\label{eq:HE3-z}
\end{eqnarray}
where the transfer-matrix is given by:\begin{widetext}
\begin{eqnarray*}
\hat{Q}_{k_{x},z}^{(3)}=\left(\begin{array}{cc}
\cos\left[p^{(3)}\left(k_{x}\right)z\right] & \frac{i\omega}{c}\frac{\varepsilon_{xx}^{(3)}}{p^{(3)}\left(k_{x}\right)}\sin\left[p^{(3)}\left(k_{x}\right)z\right]\\
\frac{ic}{\omega}\frac{p^{(3)}\left(k_{x}\right)}{\varepsilon_{xx}^{(3)}}\sin\left[p^{(3)}\left(k_{x}\right)z\right] & \cos\left[p^{(3)}\left(k_{x}\right)z\right]
\end{array}\right).
\end{eqnarray*}
\end{widetext}

In the medium $m=4$ the situation is quite different because the finite width
of this domain in $x$-direction imposes an additional boundary condition
on the slit borders $x=\pm W/2$, namely the vanishing tangential
component of the electric field $E_{z}^{(4)}\left(\pm W/2,z\right)=0$.
The solution of the Maxwell equations, satisfying these conditions%
can be expressed as follows: 
\begin{eqnarray}
\begin{aligned}
&E_{x}^{(4)}(x,z)=iW\sum_{n=0}^{\infty}\nu_{n}\cos\left[\frac{n\pi}{W}\left(x+\frac{W}{2}\right)\right]\label{eq:Ex4}\\
&\times\left\{ A_{n}^{(+)}\exp\left[i\nu_{n}\left(z-d_{\mathrm{BN}}\right)\right]\right. \\
&\left.-A_{n}^{(-)}\exp\left[-i\nu_{n}\left(z-d_{\mathrm{BN}}\right)\right]\right\} , 
\end{aligned}
\end{eqnarray}
\begin{eqnarray}
\begin{aligned}
&H_{y}^{(4)}(x,z)=\frac{i\omega\epsilon}{c}W\sum_{n=0}^{\infty}\cos\left[\frac{n\pi}{W}\left(x+\frac{W}{2}\right)\right]\label{eq:Hy4}\\
&\times\left\{ A_{n}^{(+)}\exp\left[i\nu_{n}\left(z-d_{\mathrm{BN}}\right)\right]\right.\\
&\left.+A_{n}^{(-)}\exp\left[-i\nu_{n}\left(z-d_{\mathrm{BN}}\right)\right]\right\} ,
\end{aligned}
\end{eqnarray}
\begin{eqnarray}
\begin{aligned}
&E_{z}^{(4)}(x,z)=\sum_{n=0}^{\infty}n\pi\sin\left[\frac{n\pi}{W}\left(x+\frac{W}{2}\right)\right]\\
&\times\left\{ A_{n}^{(+)}\exp\left[i\nu_{n}\left(z-d_{\mathrm{BN}}\right)\right]\right.\label{eq:Ez4}\\
&\left.+A_{n}^{(-)}\exp\left[-i\nu_{n}\left(z-d_{\mathrm{BN}}\right)\right]\right\} ,
\end{aligned}
\end{eqnarray}
where $\nu_{n}=\sqrt{\left(\frac{\omega}{c}\right)^{2}\epsilon-\left(\frac{n\pi}{W}\right)^{2}}$. 

\subsection{Boundary conditions}

The problem in course includes four boundaries between aforementioned
spatial domains, at which the fields in the neibouring domains are coupled by matching the boundary conditions. At the surface
of the upper hBN layer ($z=-d_{\mathrm{BN}}$, boundary between spatial
domains $m=1$ and $m=2$) the tangential components of the electric
and magnetic fields must be continuous across the interface, i.e.
\begin{eqnarray}
\begin{aligned}
e_{x}^{(2)}\left(k_{x},-d_{\mathrm{BN}}\right)=e_{x}^{(1)}\left(k_{x},-d_{\mathrm{BN}}\right),\label{eq:HE2md-from_HE1md}\\
h_{y}^{(2)}\left(k_{x},-d_{\mathrm{BN}}\right)=h_{y}^{(1)}\left(k_{x},-d_{\mathrm{BN}}\right).\nonumber 
\end{aligned}
\end{eqnarray}
At the interface between two hBN layers, where graphene layer is arranged
($z=0$, boundary between media $m=2$ and $m=3$) the electric field
tangential component is continuous across the interface, while the
magnetic field tangentional component is discontinuous due to presence
of two-dimensional currents $j_{x}\left(k_{x},\omega\right)$ in graphene,
\begin{eqnarray*}
\begin{aligned}
&e_{x}^{(2)}\left(k_{x},0\right)=e_{x}^{(3)}\left(k_{x},0\right),\\
&h_{y}^{(3)}\left(k_{x},0\right)-h_{y}^{(2)}\left(k_{x},0\right)=-\frac{4\pi}{c}j_{x}\left(k_{x},\omega\right).
\end{aligned}
\end{eqnarray*}
Taking into account the Ohm law, $j_{x}\left(k_{x},\omega\right)=\sigma\left(k_{x},\omega\right)e_{x}^{(2)}\left(k_{x},0\right)$
{[}where $\sigma\left(k_{x},\omega\right)$ is the conductivity of
the graphene, which in general case will be considered nonlocal, i.e.
dependent upon the in-plane wavevector $k_{x}${]}, the boundary conditions
can be expressed in the matrix form, 
\begin{eqnarray}
\left(\begin{array}{c}
h_{y}^{(3)}\left(k_{x},0\right)\\
e_{x}^{(3)}\left(k_{x},0\right)
\end{array}\right)=\hat{Q}_{g}\left(\begin{array}{c}
h_{y}^{(2)}\left(k_{x},0\right)\\
e_{x}^{(2)}\left(k_{x},0\right)
\end{array}\right)\label{eq:HE30-from-HE20}
\end{eqnarray}
with the matrix
\begin{eqnarray*}
\hat{Q}_{k_{x}}^{(g)}=\left(\begin{array}{cc}
1 & -\frac{4\pi}{c}\sigma\left(k_{x},\omega\right)\\
0 & 1
\end{array}\right).
\end{eqnarray*}

At the surfaces of the PEC film $z=d_{\mathrm{BN}}$ and $z=d_{\mathrm{BN}}+d$
(boundaries between the media $m=3,4$ and $m=4,5$, respectively) the situation is more complicated. The tangential component of the magnetic field is continuous across the interfaces over the slit area, \begin{widetext}
\begin{eqnarray}
\begin{aligned}
&H_{y}^{(3)}(x,d_{\mathrm{BN}})=H_{y}^{(4)}(x,d_{\mathrm{BN}}),\quad-\frac{W}{2}\le x\le\frac{W}{2},\label{eq:cond-Hd}\\
&H_{y}^{(5)}(x,d_{\mathrm{BN}}+d)=H_{y}^{(4)}(x,d_{\mathrm{BN}}+d),\quad-\frac{W}{2}\le x\le\frac{W}{2}.\label{eq:cond-Hdd}
\end{aligned}
\end{eqnarray}
The tangential component of the electric field has to be continuous
across the interfaces at the slit area and should vanish beyond the
slit because the metal is assumed perfect. Therefore, boundary conditions can be expressed
by the formulae
\begin{eqnarray}
\begin{aligned}
&E_{x}^{(3)}(x,d_{\mathrm{BN}})=\left\{ \begin{array}{cc}
E_{x}^{(4)}(x,d_{\mathrm{BN}}), & \quad-\frac{W}{2}\le x\le\frac{W}{2}\\
0, & \quad\mathrm{otherwise}
\end{array}\right.\label{eq:cond-Ed}\\
&E_{x}^{(5)}(x,d_{\mathrm{BN}}+d)=\left\{ \begin{array}{cc}
E_{x}^{(4)}(x,d_{\mathrm{BN}}+d), & \quad-\frac{W}{2}\le x\le\frac{W}{2}\\
0, & \quad\mathrm{otherwise}
\end{array}\right.\label{eq:cond-Edd}
\end{aligned}
\end{eqnarray}
\end{widetext}
It should be pointed out that, due to the fact that we use different bases
of eignefunctions in medium $m=4$ and in media $m=3$ and $m=5$, 
the boundary conditions \eqref{eq:cond-Hd}\textendash \eqref{eq:cond-Edd}
cannot be written in the same way as at boundaries $z=0$ and $z=-d_{\mathrm{BN}}$ where they could be expressed separately for each spatial harmonic, $e_{x}^{(m)}\left(k_{x},z\right)$ and $h_{y}^{(m)}\left(k_{x},z\right)$.
Equations \eqref{eq:cond-Hd}\textendash \eqref{eq:cond-Edd} involve the total  fields in each point $x$, $E_{x}^{(m)}(x,z)$, $H_{y}^{(m)}(x,z)$ and involve all $k_x$ harmonics. In other words, these relations are integral equations. However, they can be discretized using the specific form of the fields inside the slit, Eqs. (\ref {eq:Ex4}) and (\ref {eq:Hy4}).

\subsection{Amplitudes of the eignemodes inside the slit}

Applying consequently boundary conditions \eqref{eq:HE2md-from_HE1md}
and \eqref{eq:HE30-from-HE20} {[}jointly with expressions \eqref{eq:HE1-z-from-IR}\textendash \eqref{eq:HE3-z}
for the fields in media $m=1,2,3${]} one can obtain expressions for
the electromagnetic field tangential components of each harmonic
as
\begin{eqnarray}
\left(\begin{array}{c}
h_{y}^{(3)}\left(k_{x},d_{\mathrm{BN}}\right)\\
e_{x}^{(3)}\left(k_{x},d_{\mathrm{BN}}\right)
\end{array}\right)=\hat{F}_{k_{x}}^{(tot)}\left(\begin{array}{c}
{\cal H}_{y}^{(i)}\Theta\left(k_{c}-k_{x}\right)\\
h_{r}\left(k_{x}\right)
\end{array}\right)\label{eq:HE3d-from-IR}
\end{eqnarray}
where the total field matrix $\hat{F}_{k_{x}}^{(tot)}$ is composed from
the transfer-matrices of media $m=2,3$, boundary condition matrix
across the graphene, and the field matrix in medium $m=1$ is:
\begin{eqnarray*}
\hat{F}_{k_{x}}^{(tot)}=\hat{Q}_{k_{x},d_{\mathrm{BN}}}^{(3)}\hat{Q}_{g}\hat{Q}_{k_{x},0}^{(2)}\hat{F}_{k_{x}}^{(1)}.
\end{eqnarray*}

Substituting Eqs.\,\eqref{eq:HE5-z}, \eqref{eq:Ex4}, \eqref{eq:Hy4},  and \eqref{eq:HE3d-from-IR}
into boundary conditions \eqref{eq:cond-Hd}\textendash \eqref{eq:cond-Edd} and using orthogonality relations between the $x$ dependence of the fields in the slit, Eqs. (\ref {eq:Ex4}) and (\ref {eq:Hy4}), and the $k_x$ harmonics
(details are given in Supplementary Information), it is possible to obtain equations
for the amplitudes of forward- and backward-propagating modes inside
the slit, $A_{2n}^{(+)}$ and $A_{2n}^{(-)}$:
\begin{eqnarray}
\begin{aligned}
&\frac{W}{2}\frac{i\omega\epsilon}{c}\left(1+\delta_{n^{\prime},0}\right)\left[A_{2n^{\prime}}^{(+)}+A_{2n^{\prime}}^{(-)}\right]\label{eq:boundH3d-final}\\
&-i\frac{W^{2}}{2\pi}\sum_{n=0}^{\infty}\nu_{2n}\left[A_{2n}^{(+)}-A_{2n}^{(-)}\right] \tilde I_{2n^{\prime},2n}\left(\omega\right) \\
&=-2\frac{c}{\omega}{\cal H}_{y}^{(i)}\intop_{0}^{\infty}dk_{x}\thinspace{\cal P}_{2n^{\prime}||k_{x}}\Theta\left(k_{c}-k_{x}\right)\frac{p^{(1)}\left(k_{x}\right)}{\left[\hat{F}_{k_{x}}^{(tot)}\right]_{22}}\; ;
\end{aligned}
\end{eqnarray}
\begin{eqnarray}
\begin{aligned}
&\frac{\epsilon}{2}\left(1+\delta_{n^{\prime},0}\right)\label{eq:boundH5dd-final}\\
&\times\left[A_{2n^{\prime}}^{(+)}\exp\left(i\nu_{2n^{\prime}}d\right)+A_{2n^{\prime}}^{(-)}\exp\left(-i\nu_{2n^{\prime}}d\right)\right]\\
&-\frac{W}{2\pi}\sum_{n=0}^{\infty}\nu_{2n}I_{2n^{\prime},2n}\left(\omega\right) \\
&\times\left[A_{2n}^{(+)}\exp\left(i\nu_{2n}d\right)-A_{2n}^{(-)}\exp\left(-i\nu_{2n}d\right)\right]=0\; , 
\end{aligned}
\end{eqnarray}
where
\begin{eqnarray}
\begin{aligned}
&\mathcal{P}_{2n^{\prime}||k_{x}}=\frac{2}{W}\intop_{0}^{W/2}dx\cos\left[\frac{2n^{\prime}\pi}{W}\left(x+\frac{W}{2}\right)\right]\cos\left[k_{x}x\right]=\\
&=\frac{2}{W}\frac{k_{x}\sin\left[k_{x}\frac{W}{2}\right]}{k_{x}^{2}-\left(\frac{2n^{\prime}\pi}{W}\right)^{2}},
\label{eq:P_2n_k}
\end{aligned}
\end{eqnarray}
\begin{eqnarray*}
\tilde I_{2n^{\prime},2n}\left(\omega\right)=2\intop_{0}^{\infty}dk_{x}{\cal P}_{2n^{\prime}||k_{x}}{\cal P}_{2n||k_{x}}\frac{\left[\hat{F}_{k_{x}}^{(tot)}\right]_{12}}{\left[\hat{F}_{k_{x}}^{(tot)}\right]_{22}}
\end{eqnarray*}
and
\begin{eqnarray*}
I_{2n^{\prime},2n}\left(\omega\right)=2\intop_{0}^{\infty}dk_{x}\frac{\mathcal{P}_{2n^{\prime}||k_{x}}{\cal P}_{2n||k_{x}}}{p^{(5)}\left(k_{x}\right)}\;.
\end{eqnarray*}
Now we consider a special case where the graphene layer is deposited
directly over the slit (that is, both hBN layers are absent, i.e.
$d_{\mathrm{BN}}=0$). Then
\begin{eqnarray*}
\begin{aligned}
&\hat{F}_{k_{x}}^{(tot)}=\hat{Q}_{g}\hat{F}_{k_{x}}^{(1)}\\
&=\left(\begin{array}{cc}
1-\frac{4\pi}{\omega}\sigma\left(k_{x},\omega\right)p^{(1)}\left(k_{x}\right) & 1+\frac{4\pi}{\omega}\sigma\left(k_{x},\omega\right)p^{(1)}\left(k_{x}\right)\\
\frac{c}{\omega}p^{(1)}\left(k_{x}\right) & -\frac{c}{\omega}p^{(1)}\left(k_{x}\right)
\end{array}\right).
\end{aligned}
\end{eqnarray*}
As a consequence, Eq.\,\eqref{eq:boundH3d-final} can be rewritten
as
\begin{eqnarray}
\begin{aligned}
&\frac{W}{2}\frac{i\omega\epsilon}{c}\left(1+\delta_{n^{\prime},0}\right)\left[A_{2n^{\prime}}^{(+)}+A_{2n^{\prime}}^{(-)}\right]\\
&+i\frac{W^{2}}{2\pi}\frac{\omega}{c}\sum_{n=0}^{\infty}\nu_{2n}\left[A_{2n}^{(+)}-A_{2n}^{(-)}\right]J_{2n^{\prime},2n}\left(\omega\right)\\
&=2{\cal H}_{y}^{(i)}\intop_{0}^{\infty}dk_{x}\Theta\left(k_{c}-k_{x}\right){\cal P}_{2n^{\prime}|k_{x}},
\label{eq:boundH3dd-rewritten}
\end{aligned}
\end{eqnarray}
where
\begin{eqnarray*}
J_{2n^{\prime},2n}\left(\omega\right)=2\intop_{0}^{\infty}dk_{x}\frac{\mathcal{P}_{2n^{\prime}||k_{x}}{\cal P}_{2n||k_{x}}}{p^{(1)}\left(k_{x}\right)}\left[1+\frac{4\pi}{\omega}\sigma\left(k_{x},\omega\right)p^{(1)}\left(k_{x}\right)\right] .
\end{eqnarray*}
These equations also can be obtained from Eqs.\,\eqref{eq:boundH3d-final}
and \eqref{eq:boundH5dd-final} by taking $\hat{Q}_{k_{x}}^{(tot)}=\hat{Q}_{g}$.
Notice that the integrals $I_{2n^{\prime},2n}\left(\omega\right)$ and
$J_{2n^{\prime},2n}\left(\omega\right)$ can be calculated semi-analytically
(see Supplementary Information).

The system of equations \eqref{eq:boundH5dd-final} and  \eqref{eq:boundH3dd-rewritten} was solved by truncating it to a sufficiently large order, $n$ and checking the convergence. Once the amplitudes $A_{2n}^{(\pm)}$ have been found, the observable properties such as relectance and transmittance are calculated in a straightforward way (details are given in Supplementary Information).

\section{Suspended graphene}

\begin{figure}
\includegraphics[width=8.5cm]{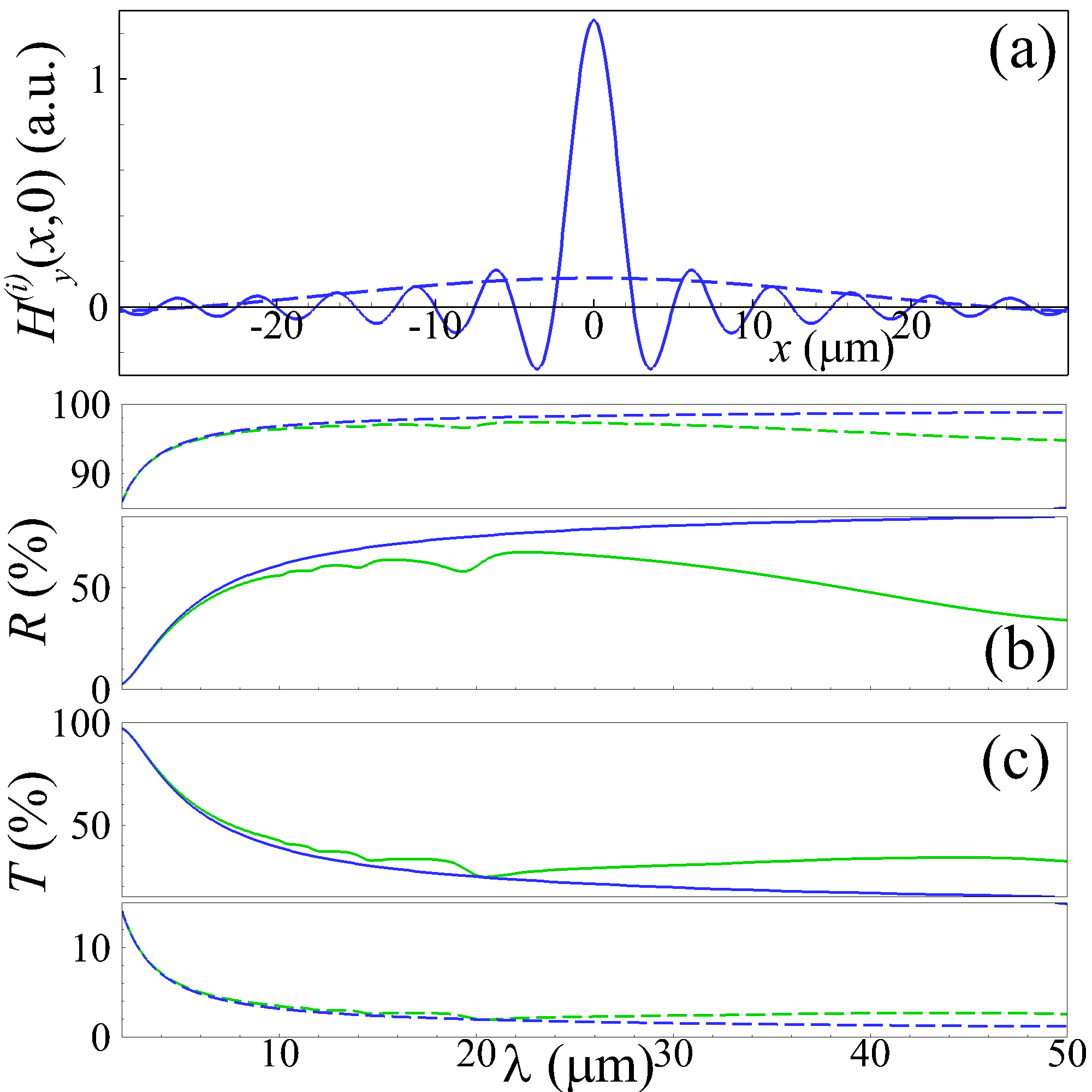}

\caption{(a) Examples of spatial shape of the incident wavepacket $H_{y}^{(i)}(x,0)$
for $\lambda=5\,\mu$m and two values of the spectral width, $ck_{c}/\omega=1$
(solid line) and $ck_{c}/\omega=0.1$ (dashed line); (b,c) Reflectance
$R$ {[}panel (b){]} and transmittance $T$ {[}panel (c){]} of graphene
monolayer suspended over the slit for two values
of the wavepacket spectral width: $ck_{c}/\omega=1$ (solid lines)
and $ck_{c}/\omega=0.1$ (dashed lines). Other parameters of the structure are: $d=100\,$nm, $W=1.5\,\mu$m, $\gamma=7.5\,$meV, $E_{F}=0\,$eV
(blue lines), $E_{F}=0.3\,$eV (green lines).}
\label{fig:RT_suspend}
\end{figure}

In order to clarify the influence of the graphene sheet on the transmittance
and reflectance of the structure we consider first the situation
where the slit is not filled ($\epsilon=1$), and both hBN
layers are absent. In other words, the graphene layer is deposited
directly on the metal film and is suspended at the area of the slit.

\begin{figure}
\includegraphics[width=8.5cm]{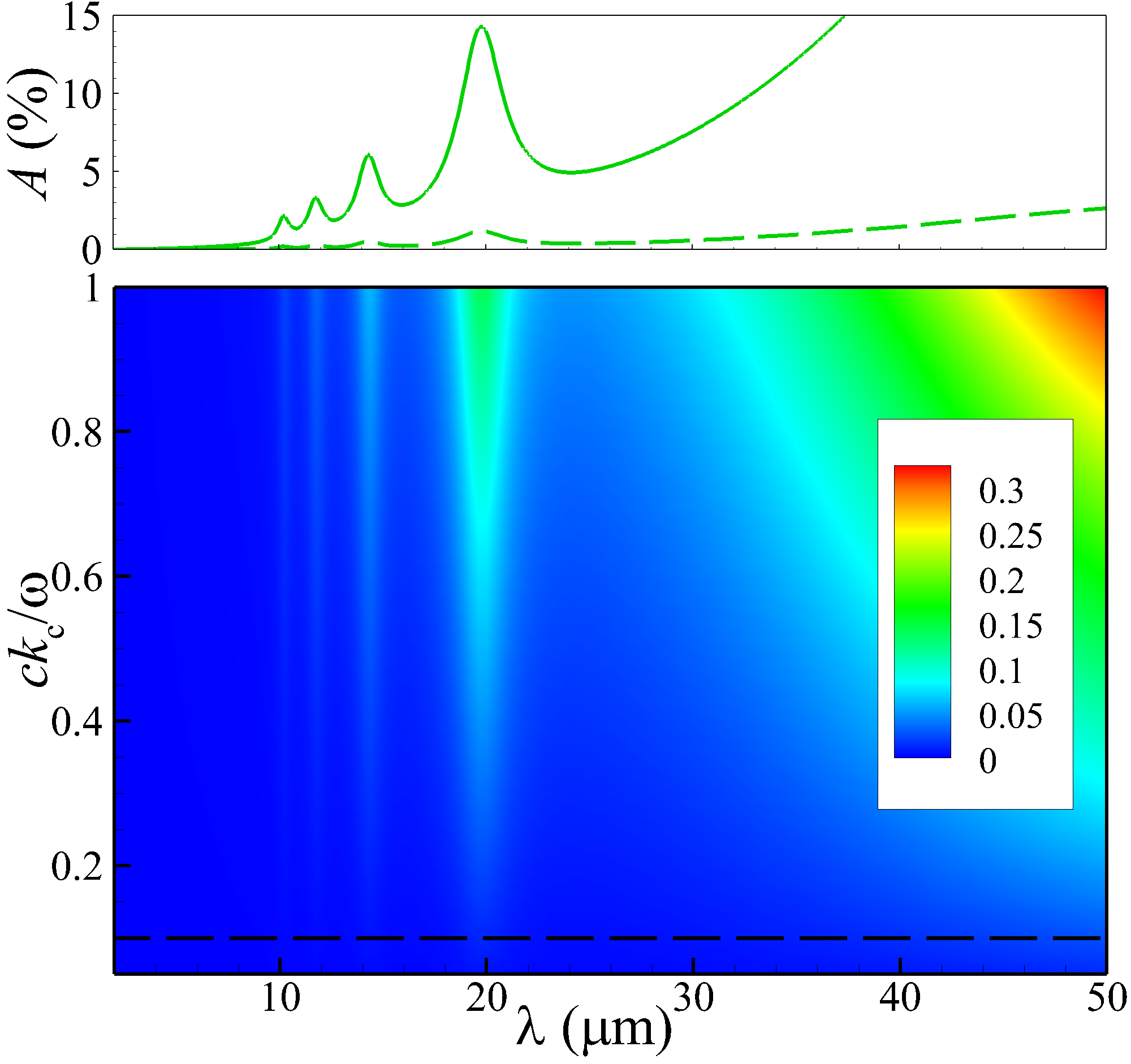}
\caption{Absorbance, $A$, {\it versus} wavelength, $\lambda$, and spectral
width of the wavepacket, $k_{c}$ (lower panel) for doped graphene with $E_{F}=0.3\,$eV suspended on top of the slit. The upper panel shows the absorbance
for two fixed walues of the spectral width, $ck_{c}/\omega=0.1$ (dashed
line, the dependence is taken along the corresponding dashed horizontal
line in the lower panel) and $ck_{c}/\omega=1$ (solid line, the dependence
is taken along the upper edge of the lower panel). The other parameters
are the same as in Fig.\,\ref{fig:RT_suspend}.}
\label{fig:A_suspend}
\end{figure}

From Fig.\,\ref{fig:RT_suspend}(a) it can be seen that a larger spectral
width $k_{c}$ (solid line) correspond to a more focused beam in the
coordinate space. As a consequence, the narrower beam (with larger $k_{c}$)
exhibits a lower reflectance, $R$, and a higher transmittance, $T$ {[}compare
dashed and solid lines in Figs.\,\ref{fig:RT_suspend}(b) and \ref{fig:RT_suspend}(c){]},
because a larger fraction of the incident wave's energy flux penetrates the slit, thus avoiding the diffraction on its edges. In
the situation of bare slit {[}$E_{F}=0\,$eV, blue lines in Figs.\,\ref{fig:RT_suspend}(b)
and \ref{fig:RT_suspend}(c){]} an increase of the wavelength, $\lambda$,
leads to the growth of the reflectance $R$ {[}see Fig.\,\ref{fig:RT_suspend}(b){]}
and decrease of the transmittance $T$ {[}see Fig.\,\ref{fig:RT_suspend}(c){]}.
This phenomenon can be accounted for the essentially
subwavelength character of the wavepacket diffraction.
In fact, in the frequency range of Fig.\,\ref{fig:RT_suspend} all the wavelengths of the wave packet are larger
than the slit width. At the same time, for a larger ratio between
the wavelength and the slit width, the presence of the slit exerts
less influence on the diffraction process, thus the reflectance becomes
more similar to that from a homogeneous PEC film, i.e. it
increases to unity with the simultaneous decrease of the transmittance.
When the slit is covered by doped graphene {[}$E_{F}=0.3\,$eV,
green lines in Figs.\,\ref{fig:RT_suspend}(b) and \ref{fig:RT_suspend}(c){]},
the aforementioned growth of the reflectance and decrease of transmittance
in nonmonotonous, demonstrating the series of local minima.
\begin{figure}
\includegraphics[width=8.5cm]{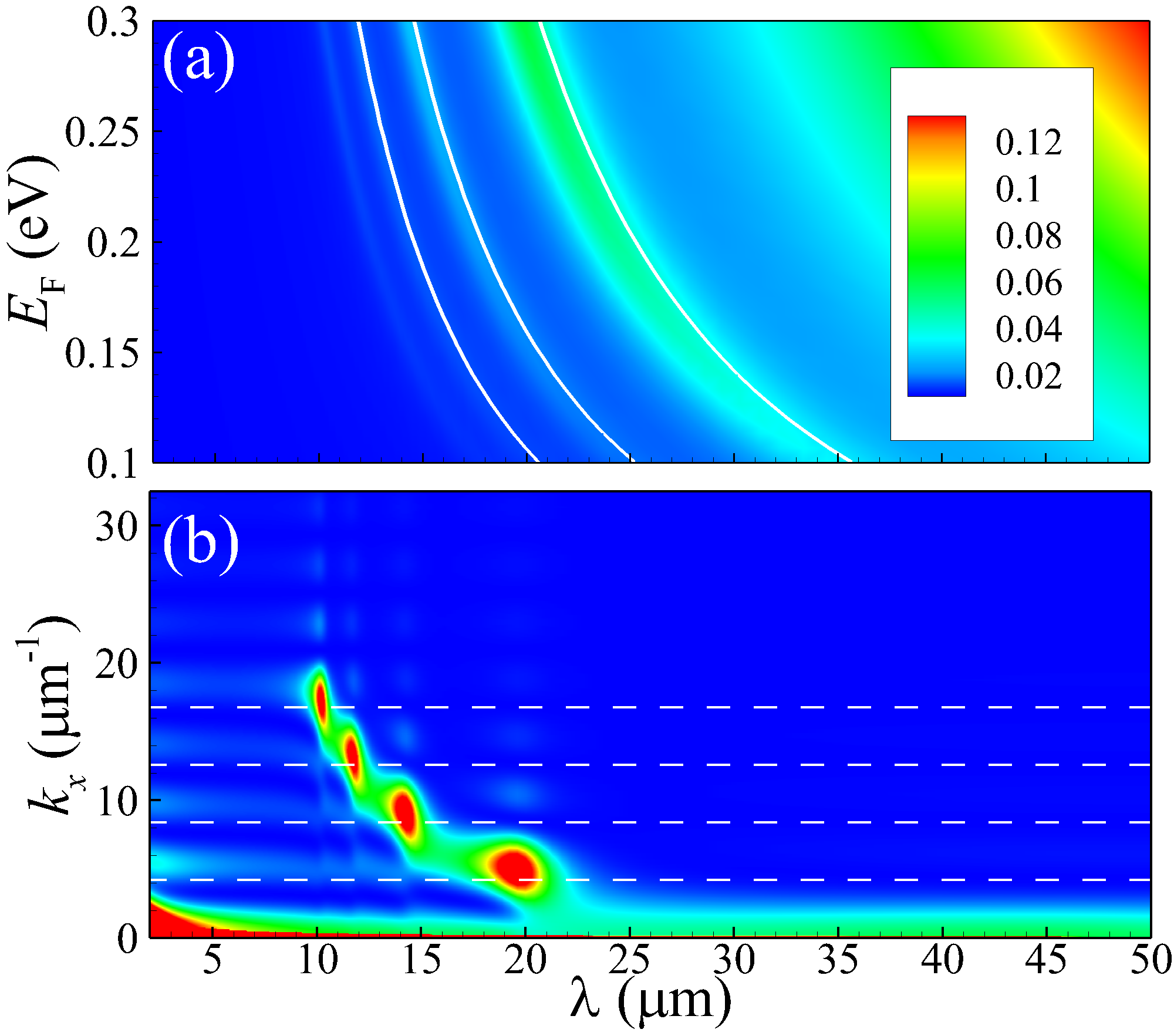}
\caption{(a) Absorbance, $A$ (depicted by color map) {\it versus} wavelength, $\lambda$,
and graphene's Fermi energy, $E_{F}$; (b) Amplitude of the electric
field of the reflected wave harmonics $\left|cp^{(1)}\left(k_{x}\right)h_{r}\left(k_{x}\right)/\omega\right|$
versus the wavelength and the wavevector, $k_{x}$, for the fixed Fermi
energy $E_{F}=0.3\,$eV. In all panels $ck_{c}/\omega=0.5$, while
other parameters are the same as in Fig.\,\ref{fig:RT_suspend}.
Solid white lines {[}superimposed on the color map in panel (a){]} demonstrate
the SPP eigenmodes with the wavevectors $k_{x}=2n\pi/W$ ($n=1,2,3$,
from right to left), the same wavevectors are depicted in panel (b)
by white dashed horizontal lines.}
\label{fig:A_Ef_suspend}
\end{figure}

Quite interesting are the absorption spectra of the considered structure.
As can be seen from the upper panel of Fig.\,\ref{fig:A_suspend}, the absorbance, $A=1-R-T$, is high at the wavelengths corresponding
to the reflectance and transmittance minima. Furtermore, a larger spectral
width of the incident wavepacket ($k_{c}$) makes these absorbance peaks
more pronounced {[}see lower panel of Fig.\,\ref{fig:A_suspend}{]},
while their positions (wavelengths $\lambda$) are not affected. At the same time, the positions of the absorption peaks are strongly influenced by the graphene's Fermi
energy {[}see Fig.\,\ref{fig:A_Ef_suspend}(a){]}. This
fact resembles the crucial property of graphene SPPs whose dispersion curve, $\omega (k_{x})$, scales with the Fermi energy approximately as $\omega \propto \sqrt {k_xE_F}$ for small $k_x$. A more detailed analysis leads to the following expression for the characteristic wavelengths of the SPP eigenmodes:
\begin{eqnarray}
\begin{aligned}
&\lambda_{n}=\frac {\sqrt 2\pi \hbar c}{\alpha E_{F}} \\
&\times\left(\sqrt{1+\left(\frac{2n\pi\hbar c}{2\alpha E_{F}W}\right)^{2}}-1\right)^{- 1/2},\label{eq:lambda_n}
\end{aligned}
\end{eqnarray}
where $n=1, 2, 3...$ and $\alpha $ stands for the fine structure constant. These modes are depicted the white solid lines in Fig.\,\ref{fig:A_Ef_suspend}(a). The polaritonic
character of the absorption peaks are confirmed by the fact that the spectral 
positions of the absorbance maxima coincide with the graphene SPP eigenmodes. Moreover, the modulus of the electric field of the reflected
wave harmonics {[}depicted in Fig.\,\ref{fig:A_Ef_suspend}(b){]}
has its maxima near the resonance wavelengths,
$\lambda_{n}$ {[}Eq.\,\eqref{eq:lambda_n}{]}. The associated wavevectors $k_{x}=2\pi n/W$
{[}horizonal dashed lines in Fig.\,\ref{fig:A_Ef_suspend}(b){]} correspond to the even spatial profiles of the tangential
components of the  eigenmode's electromagnetic field inside the slit {[}see Eqs.\,\eqref{eq:Ex4},
\eqref{eq:Hy4}{]}. In other words, when the incident wave packet's
wavelength coincides with the wavelength of an SPP eigenmode with the wavevector
$k_{x}=2\pi n/W$, the wave packet, being diffracted on the slit, resonantly
excites SPPs in graphene. This polariton, owing to the multiple reflections from
the slit edges, forms the SPP standing wave in the suspended graphene
with the nodes of its electric field ($x$-component) at the edges of
the slit. This resonant excitation leads to the transformation of incident
wave packet's energy into the energy of the with $E_{F}=0.3\,$eV standing wave; this phenomenon gives rise to the resonant absorption.
 
\begin{figure}
\includegraphics[width=8.5cm]{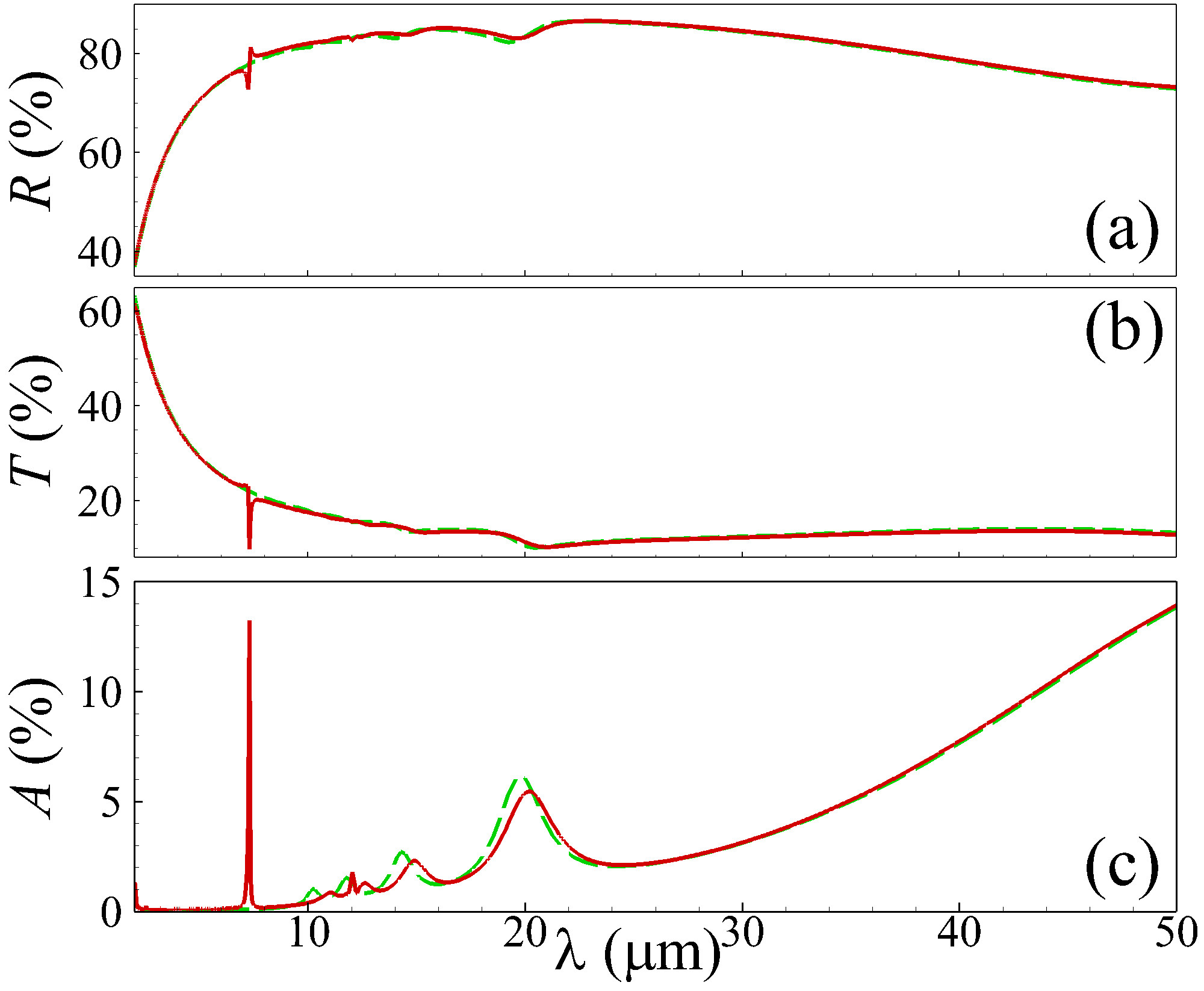}
\caption{Reflectance {[}panel (a){]}, transmittance {[}panel (b){]}, and absorbance
{[}panel (c){]} of (i) suspended graphene layer 
{[}dashed green lines{]} and (ii) graphene layer cladded by two hBN layers with thicknesses $d_{\mathrm{BN}}=1\,$nm.
In both cases with $E_{F}=0.3\,$eV and the incident wavepacket is characterized by the spectral width $ck_{c}/\omega=0.5$. Other parameters are the same as
in Fig.\,\eqref{fig:RT_suspend}.}
\label{fig:RT_hBN}
\end{figure}
Qualitatively, the situation here is similar to the structure composed of non-absorbing nanoparticles (NPs) deposited on a homogeneous graphene sheet, where the symmetry breaking caused by the NPs induces surface plasmon-polaritons and originates absorption of propagating EM waves due to energy dissipation by the graphene plasmons \cite {Santos2014}. Indeed, the absorbance spectra of Fig. \,\ref{fig:RT_suspend} show a similarity to that of the "graphene + NPs" system \cite {Santos2014}, although here we observe not just the lowest energy (longest wavelength) absorption peak but also its overtones according to Eq. \eqref {eq:lambda_n}. 

\section{Effect of the substrate}

The essential physics behind the optical properties of the graphene-covered slit has been established in the previous section and now we may address a further question, how will the results change if the
graphene is not deposited directly on the metal film but rather cladded
by two hBN layers as depicted in Fig. 1? To answer this question, in Fig.\,\ref{fig:RT_hBN} we present a comparison of the reflectance, transmittance, and absorbance of
the suspended (dashed green lines) and  hBN-encapsulated graphene (solid red lines). As can be seen from the
comparison of the dashed and solid lines in Fig.\,\ref{fig:RT_hBN},
the presence of hBN leads to the small shift of the positions of the SPP
absorbance maxima, and the appearance of two additional maxima (nearby
$\lambda\approx7.3\thinspace\mu$m and $\lambda\approx12\thinspace\mu$m)
due to the excitation of optical phonons in the hBN cladding layers.
\begin{figure}
\includegraphics[width=8.5cm]{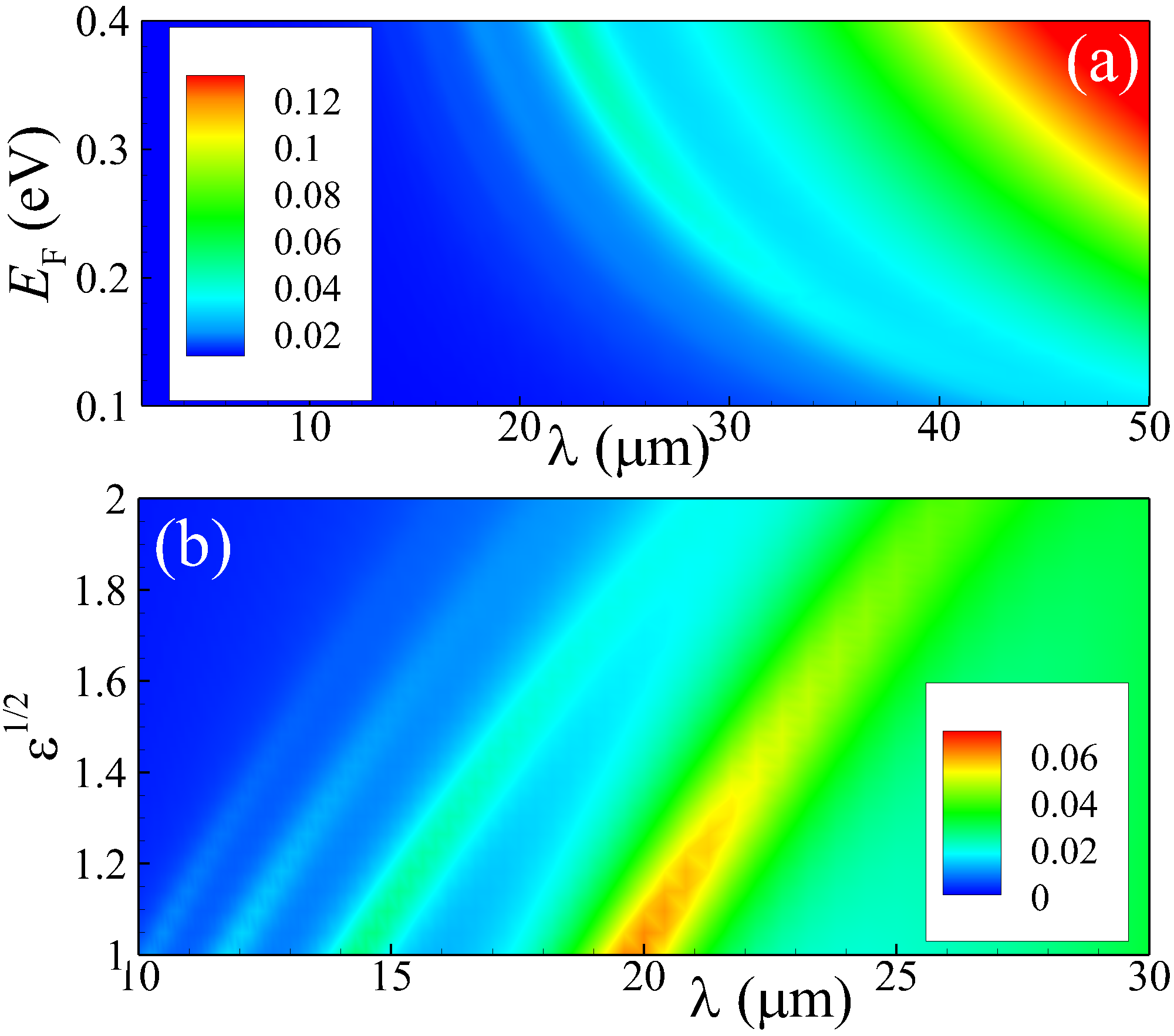}
\caption{(a) Absorbance, $A$, {\it versus} wavelength, $\lambda$, and Fermi energy,
$E_{F}$, for the slit filled by a dielectric with $\epsilon=3.9$; (b) Absorbance {\it versus} waveleng and refractive index, $\epsilon^{1/2}$, for the graphene Fermi
energy $E_{F}=0.3\,$eV. In both panels $ck_{c}/\omega=0.5$ and all other
parameters are the same as in Fig.\,\ref{fig:RT_suspend}.}
\label{fig:A_epsilon}
\end{figure}

One of the crucial properties of SPPs is the strong sensitivity of their
dispersion to the dielectric constants of the surrounding media, which should also apply to the dielectric  filling the slit in the present case.
Therefore, Fig.\,\ref{fig:A_epsilon}(a) presents the
absorbance of the graphene layer deposited over the slit filled by a material with the dielectric constant  $\epsilon=3.9$. As it follows from the
comparison of Figs.\,\ref{fig:A_epsilon}(a) and \ref{fig:A_Ef_suspend}(a),
the presence of the dielectric inside the slit results in a red shift
of the plasmonic absorption peaks. In more detail this phenomenon
is demonstrated in Fig.\,\ref{fig:A_epsilon}(b). As it can be seen from this plot,
there is an almost linear dependence of the plasmonic absorption peak
positions upon the refractive index, $\epsilon^{1/2}$. Even more,
when the refractive index is changed in the limits between 1 and 2,
the plasmonic absorption peak wavelength is shifted by $\sim5\,\mu$m. 

This phenomenon can be used in plasmonic sensors. The advantage
of the graphene-based plasmonic sensor is an additional degree of its
dynamical tunability. Thus, if a source of electromagnetic readiation
with a fixed wavelength is used, then the position of the plasmonic absorption 
peak can be tuned to the desired wavelength by adjusting the graphene's
Fermi energy and its value can provide 
information about the dielectric constant of the medium that fills
the slit.

\vspace{0.4cm}
\section{Conclusion}

To conclude, we considered the diffraction of the spatially localized wavepacket
on the single slit in a perfect metalic film covered by monolayer graphene. We have shown that this geometry is suitable for the excitation of surface plasmon-polaritons in graphene.
The diffraction of the wavepacket on the slit is accompanied by the excitation of the polariton standing wave, for which the vertical edges of the slit in PEC serve as a cavity. 
The resonance condition for the excitation of such standing waves can be
expressed in the following manner: the excitation of SPP takes place for a given frequency of the incident wavepacket, $\omega $, if the slit width contains an integer number of the polariton wavelengths, $\lambda =2\pi /k_x$, where $k_x$ is the SPP wavevector for the frequency $\omega $. 

The excitation of SPPs is manifested by the appearance of the peaks in the
absorbance spectrum. Also the positions of these resonant absorption peaks
can be effectively tuned by the electrostatic gating of graphene.
The resonant frequencies are shown to be very sensitive to the refractive
index of the medium, which fills the slit. This phenomenon can be used for environment sensing. The advantage of the such graphene-based sensor
is the possibility to tune (by the graphene gating) the position of the absorption peaks to the spectral range where the fingerprints of the molecules are the most intense, e.g. due to the presence of dipolar vibration modes.

\section*{Acknowledgements}

Authors acknowledge support from the European Commission through the
project ''Graphene- Driven Revolutions in ICT and Beyond''
(Ref. No. 785219), and the Portuguese Foundation for Science and Technology through the Strategic Funding UID/FIS/04650/2019.
Additionally, authors acknowledge financing from FEDER and the portuguese
Foundation for Science and Technology (FCT) through project POCI-01-0145-FEDER-028114. 

\bibliographystyle{unsrt}
\bibliography{difr_slit_bibtex}

\end{document}